\documentstyle{mn}

\input epsf

\begin{document}
\journal{Preprint astro-ph/0011384}
\title[Apparent and actual galaxy cluster temperatures]
{Apparent and actual galaxy cluster temperatures}
\author[A.~R.~Liddle, P.~T.~P.~Viana, A.~K.~Romer and R.~G.~Mann]{Andrew 
R.~Liddle$^1$, Pedro T. P. Viana$^{2}$, A. Kathy Romer$^3$ and Robert G. 
Mann$^4$\\
$^1$Astronomy Centre, University of Sussex, Falmer, Brighton BN1 9QJ, UK\\
$^2$Centro de Astrof\'{\i}sica da Universidade do Porto,
Rua das Estrelas s/c, 4100 Porto, Portugal\\
$^3$Physics Department, Carnegie Mellon University, Pittsburgh PA15213, U.S.A.\\
$^4$Institute for Astronomy, University of Edinburgh, Blackford Hill, Edinburgh, 
Eh9 9HJ}
\maketitle
\begin{abstract}
The redshift evolution of the galaxy cluster temperature function is a
powerful probe of cosmology. However, its determination requires the
measurement of redshifts for all clusters in a catalogue, which is
likely to prove challenging for large catalogues expected from
XMM--Newton, which may contain of order 2\,000 clusters with measurable 
temperatures distributed
around the sky. In this paper we study the {\em apparent} cluster
temperature, which can be obtained without cluster redshifts. We show
that the apparent temperature function itself is of limited use in
constraining cosmology, and so concentrate our focus on studying how
apparent temperatures can be combined with other X-ray information to
constrain the redshift. We also briefly study the
circumstances in which non-thermal spectral features can give
redshift information.
\end{abstract}

\section{Introduction}

Considerable attention has been devoted to the study of the evolution
of the galaxy cluster temperature function with redshift, which
promises to be an extremely powerful probe of the density
parameter (Frenk et al.~1990; Oukbir \& Blanchard 1992; Viana \& Liddle 1996; 
Eke, Cole \& Frenk 1996). 
Even very small numbers of high-mass, high-redshift
clusters can rule out the critical-density paradigm; indeed, several
authors claim that they have already done so (Henry 1997; Bahcall \& Fan 1998;  
Eke et al.~1998) though this remains
controversial (Sadat, Blanchard \& 
Oukbir 1998; Reichart et al.~1999; Viana \& Liddle 1999). In order to fully 
apply this method, the cluster masses
must be accurately determined and the usual technique is to use the
gas temperature as measured from the X-ray emission. In addition the
cluster redshift is required, both to place it correctly in the
evolutionary sequence, and because the redshift is needed to convert
the apparent temperature into the actual cluster temperature.

For existing catalogues of clusters for which the temperatures could
be estimated, obtaining the redshifts proved a manageable task, as the
number of clusters with sufficient photon counts to allow temperature
determination was small. This is set to change with observations
by the XMM--Newton (hereafter just XMM) satellite; in a recent paper
(Romer et al.~1999) we showed that a planned serendipitous cluster
survey which will analyze all XMM--EPIC frames suitable for serendipitous 
cluster detection, {\sc Xcs}\footnote{See {\tt
www.xcs-home.org}\ \ for further details.}, may contain as many as
10\,000 galaxy clusters of temperature 2 keV and above, of which
around 2\,000 may have sufficient photon counts to allow the
temperatures to be accurately estimated without further X-ray observations.  
Given that these will be
distributed more or less randomly across the sky, follow-up to
obtain spectroscopic redshifts represents a substantial task.  The
main focus of this paper is on the use of the full available X-ray
information to optimize the follow-up efficiency onto the
high-redshift population.

Although a survey like the {\sc Xcs} is likely to contain around
two thousand clusters with high enough photon counts to permit an
accurate temperature estimate, the thermal
bremsstrahlung spectrum only gives the apparent temperature
\begin{equation}
T_{{\rm app}} = \frac{T}{1+z} \,,
\end{equation}
with the true temperature not being known until the redshift is determined.
For very luminous clusters this degeneracy may be broken by visible spectral 
lines such as the Iron K line complex at $7$ keV, but this will be challenging 
for most 
clusters and we defer discussion of this possibility until the end of the paper.

In this paper we discuss several aspects of apparent temperatures and the 
estimation 
of cluster redshifts from X-ray data. Apparent temperatures of clusters were 
first discussed by Oukbir \& Blanchard (1997) in the context of the ROSAT 
all-sky survey. They also noted the curious point that even in the absence of 
redshifts the apparent temperature should still be a
good estimator of relative cluster masses;
for example in a critical-density model scaling laws predict $M
\propto T^{3/2}/(1+z)^{3/2} \propto T_{{\rm app}}^{3/2}$. 

We focus on issues of apparent temperatures relevant to the XMM satellite.
First of all, we analyze whether the apparent temperature function (that is, the
number density of clusters observed above a given apparent temperature $T_{{\rm
app}}$) might in itself prove a useful probe of cosmology.  The answer will be
that it proves of limited use, demonstrating the importance of determining
cluster redshifts at the earliest possible stage.  In that light, we go on to
consider how other X-ray observables can be combined with the apparent
temperatures in order to constrain the redshifts, particularly with a view to
eliminating low-redshift clusters from the follow-up candidate list.  The main
observables are the angular size and the apparent luminosity of the clusters,
and it is the latter which proves powerful in combination with the apparent
temperature.  We also briefly study {\sc xspec} spectral simulations to assess
the likelihood of redshift determination from X-ray spectral lines.

\section{The apparent temperature function}

Before considering redshift estimation, it is worth
exploring whether the X-ray apparent temperature function $N(>T_{{\rm
app}})$ might yield useful constraints on cosmology. Unlike the real
temperature function $N(>T,z)$, which can also be taken as a function
of redshift, the apparent temperature function includes clusters from
all redshifts. The main application of the cluster number density is
to limit the matter density in the Universe, and so we focus on that.

\begin{figure}
\centering
\leavevmode\epsfysize=5.6cm \epsfbox{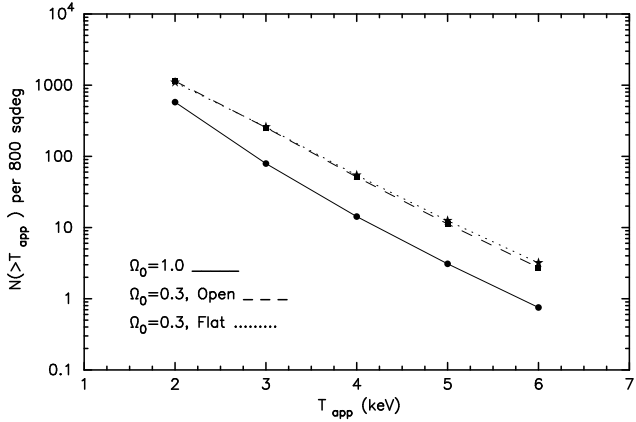}\\
\caption[Fig1]{The number of clusters with apparent temperature above
a given value, plotted for three different cosmologies. The numbers
assume a survey area of 800 square degrees, and assume a catalogue
with observed flux limit $10^{-13} \, {\rm erg} \, {\rm s}^{-1} \,
{\rm cm}^{-2}$.}
\end{figure}

We compute the apparent temperature function using the
Press--Schechter techniques of Viana \& Liddle (1999), to which we
refer the reader for details.\footnote{For an up-to-date analysis of cluster 
abundance constraints including corrections to Press--Schechter at the high-mass 
end, see Pierpaoli, Scott \& White (2000).} The key assumptions of the method 
are
that the temperature can be obtained from the mass via the usual
scaling relations (normalized to hydrodynamical cluster simulations),
and that the relationship between luminosity and temperature observed
in the present Universe (Allen \& Fabian 1998) is valid also at high redshift. 
This latter
assumption is quite likely to prove incorrect at some level and is
subject to modification when improved observations become
available. We study the same three cold dark matter (CDM) cosmologies
as in Viana \& Liddle (1999). One is a critical-density cosmology, and
the other two are low-density models with $\Omega_0=0.3$, one of which
is an open model and the other the currently-favoured spatially-flat
model with a cosmological constant. In each case the models are
normalized to give a good fit to the present-day cluster number
density (Viana \& Liddle 1996, 1999) by adjusting the dispersion
$\sigma_8$ of the power spectrum.

\begin{figure}
\centering
\leavevmode\epsfysize=5.6cm \epsfbox{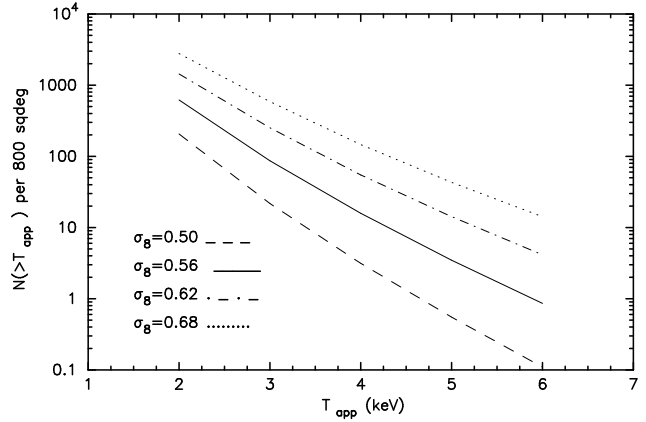}\\
\caption[Nabove]{The apparent temperature function for
critical-density models with different $\sigma_8$ as shown. The curve
$\sigma_8 = 0.56$ is the standard cluster normalization used in
Figure~1.}
\end{figure}

Figure~1 shows the apparent temperature function predicted, assuming a
survey with a flux limit of $10^{-13} \, {\rm erg} \, {\rm s}^{-1} \,
{\rm cm}^{-2}$ in the [0.5, 2] keV band, which is around the mean
level at which XMM would expect to have sufficient photons for
temperature estimation in a typical pointing. The curves look quite
promising, with a factor of a few difference between the low-density
and critical-density cases. Unfortunately though, this does not take
into account the effect of varying other parameters, and it transpires
that there is a strong degeneracy with the normalization $\sigma_8$ of
the matter power spectrum. Figure~2 shows the predictions for a series
of critical-density models with different values of $\sigma_8$, and
shows a strong dependence which is capable of swamping the dependence
on $\Omega_0$. Taking for example the flat case, the normalization of
the power spectrum we use takes $\sigma_8 \Omega_0^{-0.47} = 0.56$
(Viana \& Liddle 1999), with an uncertainty of about 20 per cent at the 95\%
confidence level, so all the curves shown in Figure~2 are
plausible. From the curves, we estimate that the number of clusters
above a given apparent temperature scales roughly as
\begin{equation}
N \propto \left(\sigma_8 \Omega_0^{-0.47}\right)^8 \, \frac{1}{\Omega_0} \,,
\end{equation}
so the first dependence dominates the latter. This is the familiar
degeneracy of the low-redshift cluster sample, arising because at any given 
apparent temperature the sample is dominated by nearby clusters. The
apparent temperature function can therefore be used to measure that combination
to high accuracy, but nothing else.

Although the previous figures illustrate the expected behaviour, the
assumption of a fixed flux cut is over-simplistic for XMM. In
reality, whether or not a cluster has a measurable temperature depends
on the cluster temperatures and redshifts themselves, and also on the
distribution of pointing durations. In Romer et al.~(1999) we
established a large set of simulations allowing us to determine which
clusters can be identified in XMM frames, and the subset of these for
which temperature estimation is available. Of course, given those data
a fixed flux cut-off can be imposed to give a flux-limited
sample, but we can also simulate the complete expected
data-set. We have done so, and we have found it makes negligible
difference to the predicted curves.

\begin{figure}
\centering
\leavevmode\epsfysize=5.3cm \epsfbox{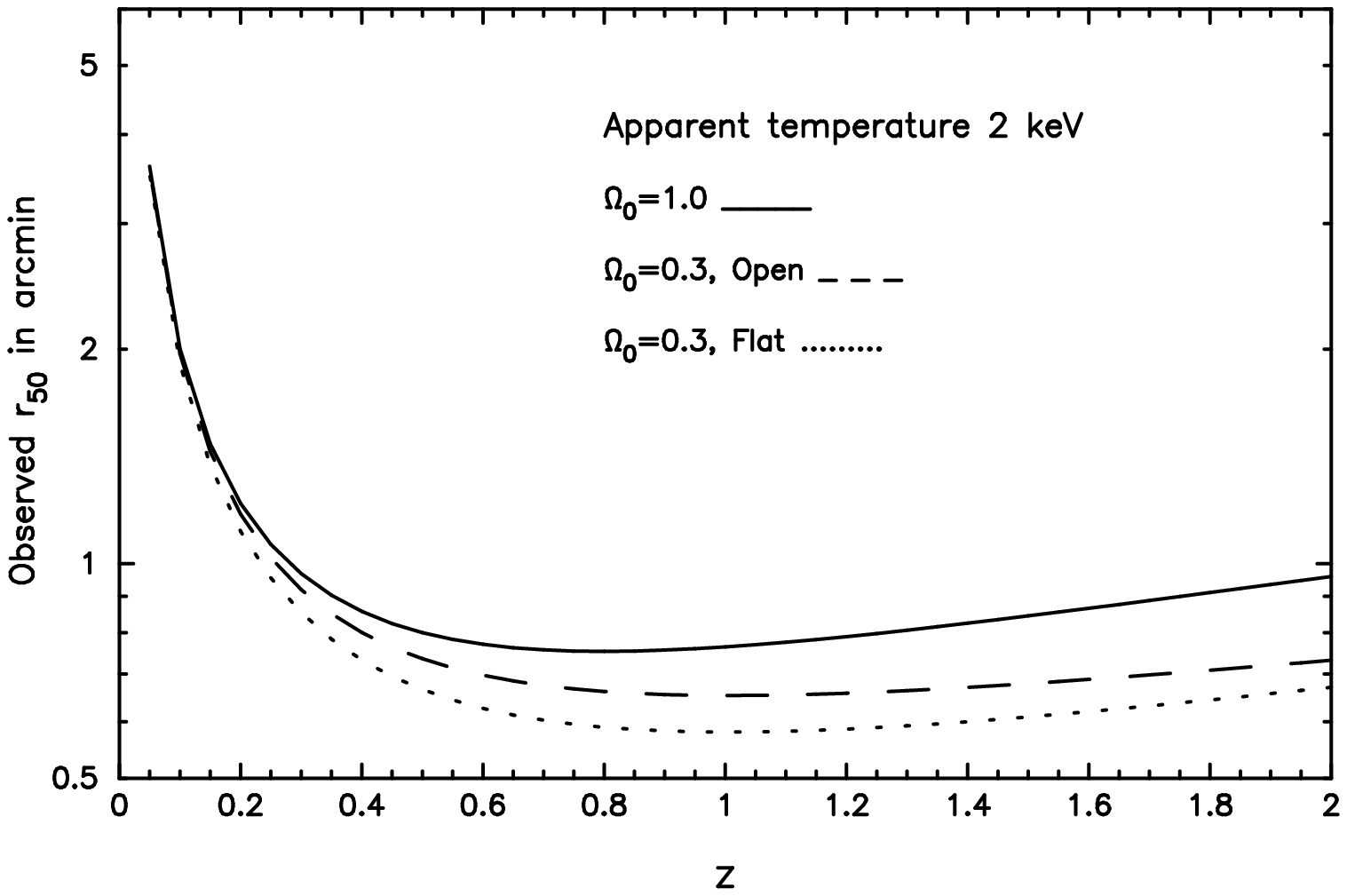}\\
\leavevmode\epsfysize=5.3cm \epsfbox{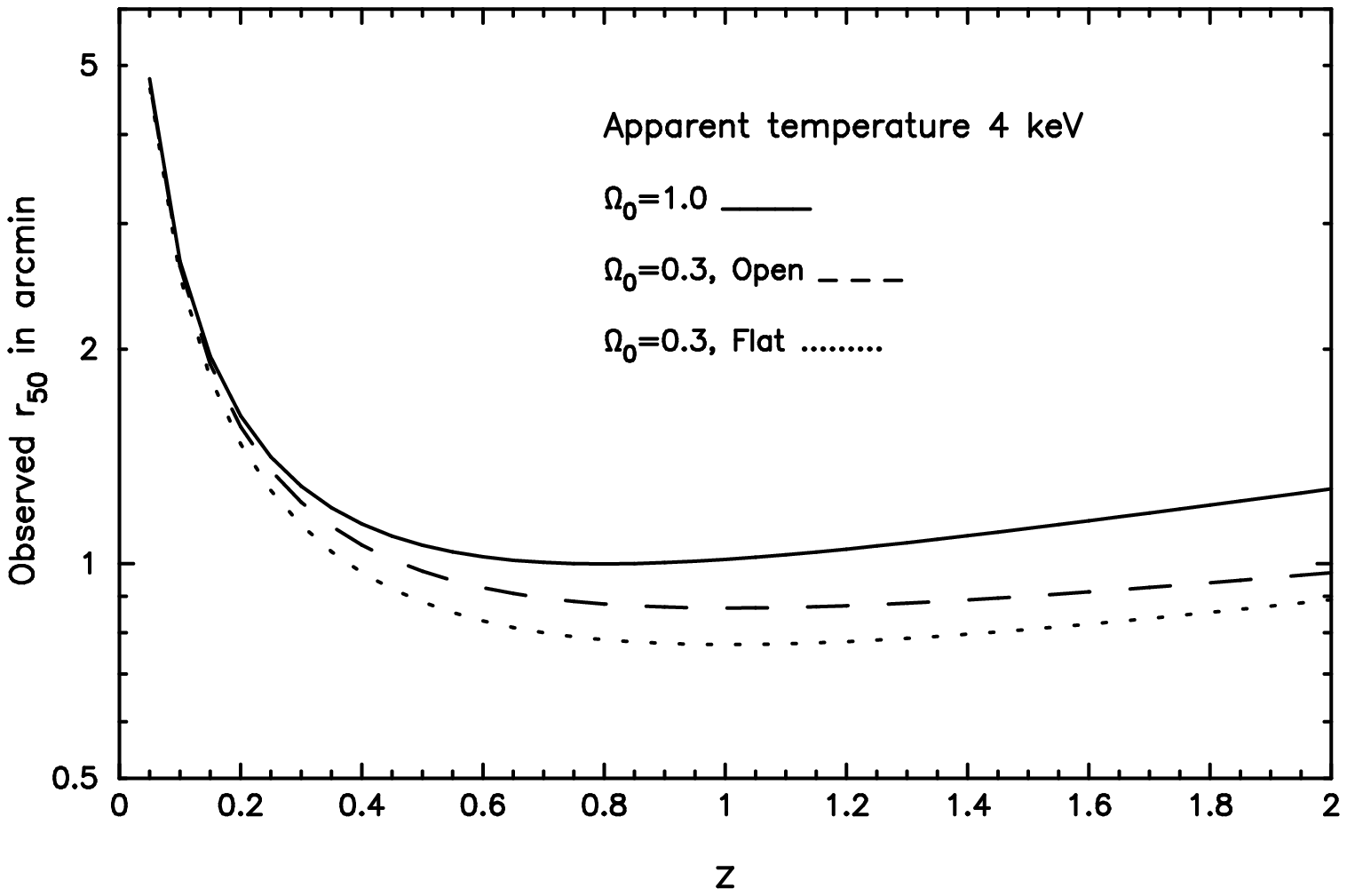}\\
\caption[Ang2]{The predicted angular extent as a function of redshift for a 
clusters of apparent temperatures 2 keV (top) and 4 keV (bottom).}
\end{figure}

\section{Redshift estimates from X-ray observables}

In order to fully capitalize on a large X-ray cluster catalogue,
redshifts are clearly essential for those clusters with measured
apparent temperatures. In cosmology, this allows the evolution of the 
temperature
function to be used to break the low-redshift degeneracy. Given the expected
size of the data set, full spectroscopic follow-up is a substantial
task, and in this section we consider ways in which other X-ray observables can 
be used along with the apparent temperatures
to help optimize the follow-up strategy by providing estimated cluster
redshifts.

\begin{figure}
\centering
\leavevmode\epsfysize=5.3cm \epsfbox{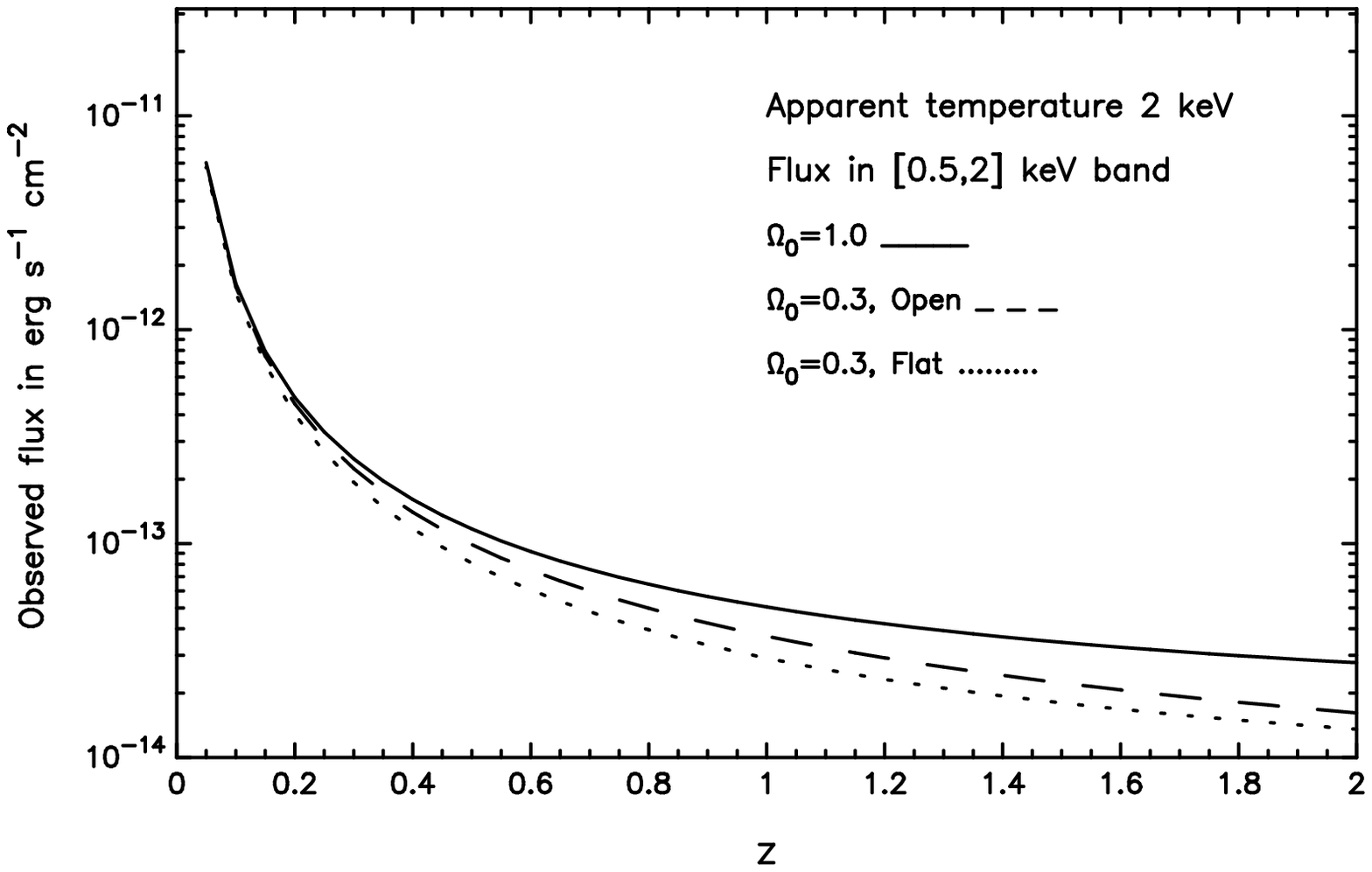}\\
\leavevmode\epsfysize=5.3cm \epsfbox{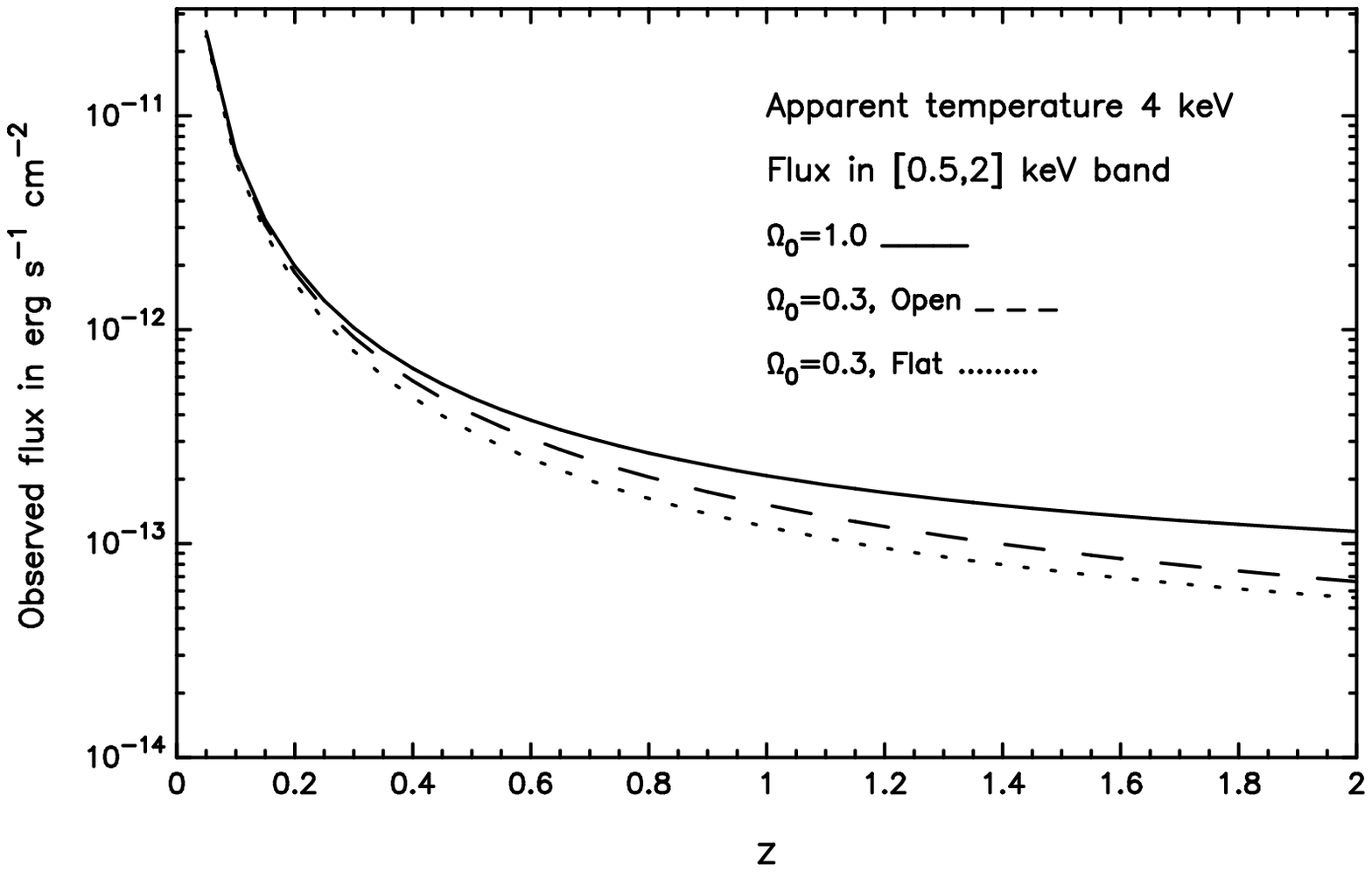}\\
\caption[Flux2]{The predicted observed flux as a function of redshift
for a clusters of apparent temperatures 2 keV (top) and 4 keV
(bottom). We have followed the convention of giving the flux in the 
ROSAT band, though the band is irrelevant as clusters of the same apparent 
temperature share the same spectral shape.}
\end{figure}

\subsection{X-ray flux and angular extent}

In addition to the spectrum of the photons received, the main X-ray
observables are the angular extent of the clusters and the observed
flux.  In Figures~3 and 4 we plot the expected redshift dependence of
these, for the three cosmologies we are considering, and for clusters
of apparent temperature 2 keV and 4 keV.  Note that these plots
differ from more standard ones in that it is the apparent temperature
which is fixed, so that as the cluster is moved to higher redshift its
temperature increases.  The quantity $r_{50}$ is the radius of the
region enclosing the inner 50 per cent of the total cluster flux,
which for a cluster with an isothermal $\beta$-profile and $\beta=2/3$
is a factor of $\sqrt{3}$ larger than the cluster core radius,
$r_{{\rm c}}$.  Unfortunately, at present the
mechanism that gives rise to the cluster core radius is not well understood, and 
hence neither is how
$r_{{\rm c}}$ is related to X-ray temperature or luminosity nor how it may
change with redshift. Therefore we simply take the empirical relation
between $r_{{\rm c}}$ and X-ray luminosity given in Jones et al.~(1998) and
assume it does not evolve with redshift. Luminosity, also, cannot yet
be predicted from first principles, and so we assume that the observed
low-redshift relation between temperature and luminosity does not
evolve with redshift.  Once XMM has observations to high redshift, any
evolution is readily incorporated.

These plots confirm the standard beliefs, based on consideration of clusters of
fixed properties viewed at different redshifts, that the apparent size is only a
weak function of redshift beyond 0.3 or so,\footnote{The size may however be a
very good estimator of apparent temperature (see e.g.~Mohr et al.~2000), and the 
size--temperature relation may prove a useful probe of cosmology (Verde et 
al.~2000).}  while
the flux continues to evolve significantly.  The former is therefore not useful
as a redshift discriminator, even though XMM should resolve all clusters, while
the latter is.  We checked whether the situation would change if the core radius
evolved in a self-similar way, maintaining a fixed size relative to the virial
radius, but though in this case the apparent size does depend more on redshift,
it still does not change as strongly as the flux.

In order to ascertain how useful the flux is as a redshift estimator, and 
to get a feeling for the importance of scatter in the luminosity--temperature 
relation and uncertainties in the measured temperatures, we have carried out 
Monte Carlo simulations of catalogues corresponding to three years of {\sc 
Xcs} data. We assume a flat, cosmological constant dominated, Universe with
$\Omega_{0}=0.3$. A realistic simulation needs to include both intrinsic scatter 
in the luminosity at a given temperature, and the errors in temperature 
measurement. The intrinsic scatter in the luminosity 
distribution is estimated from hydrodynamical simulations as being 20 per 
cent at one-sigma (O.~Muanwong, private communication). 

\begin{figure}
\centering
\leavevmode\epsfysize=6cm \epsfbox{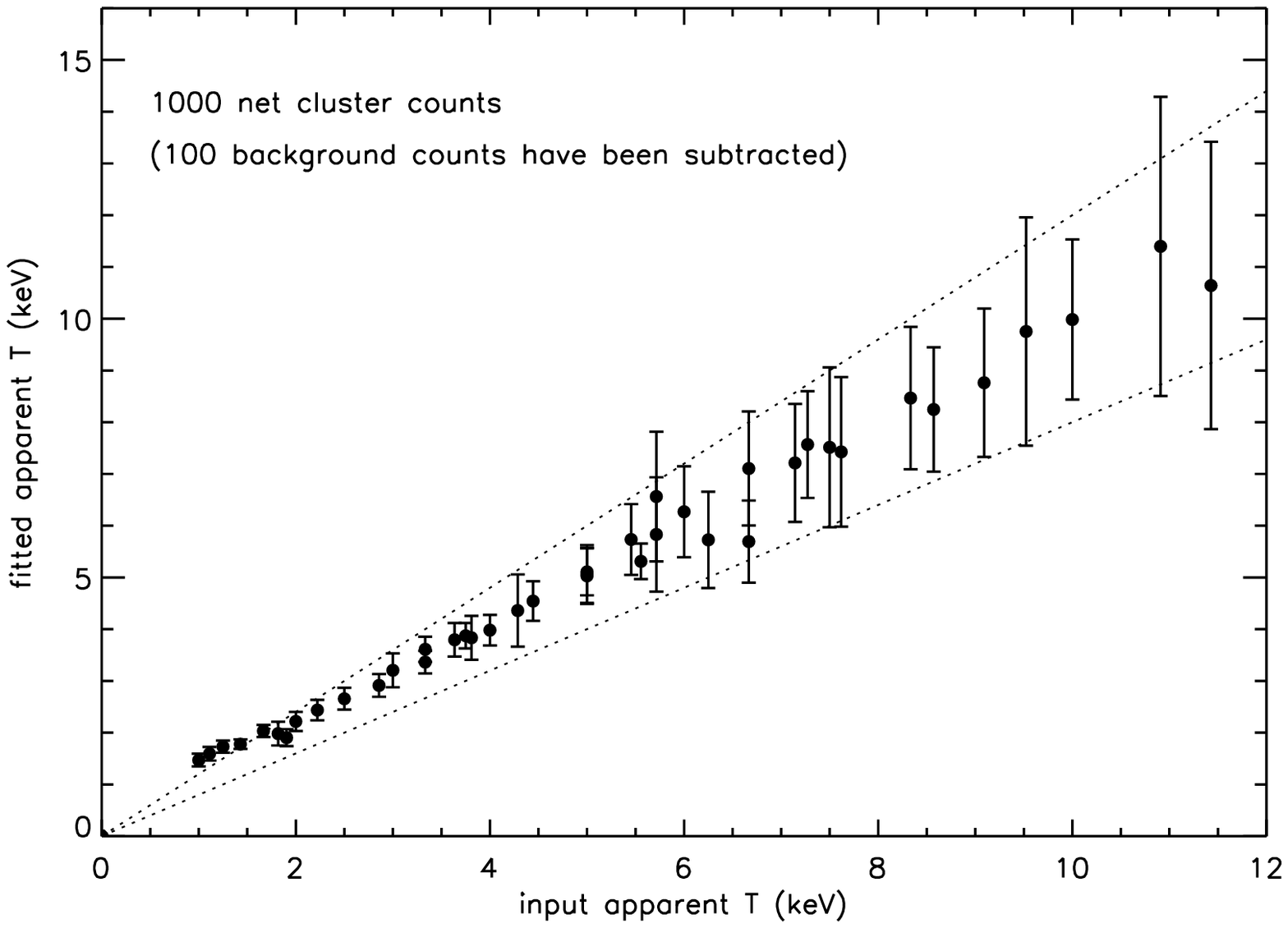}\\
\leavevmode\epsfysize=6cm \epsfbox{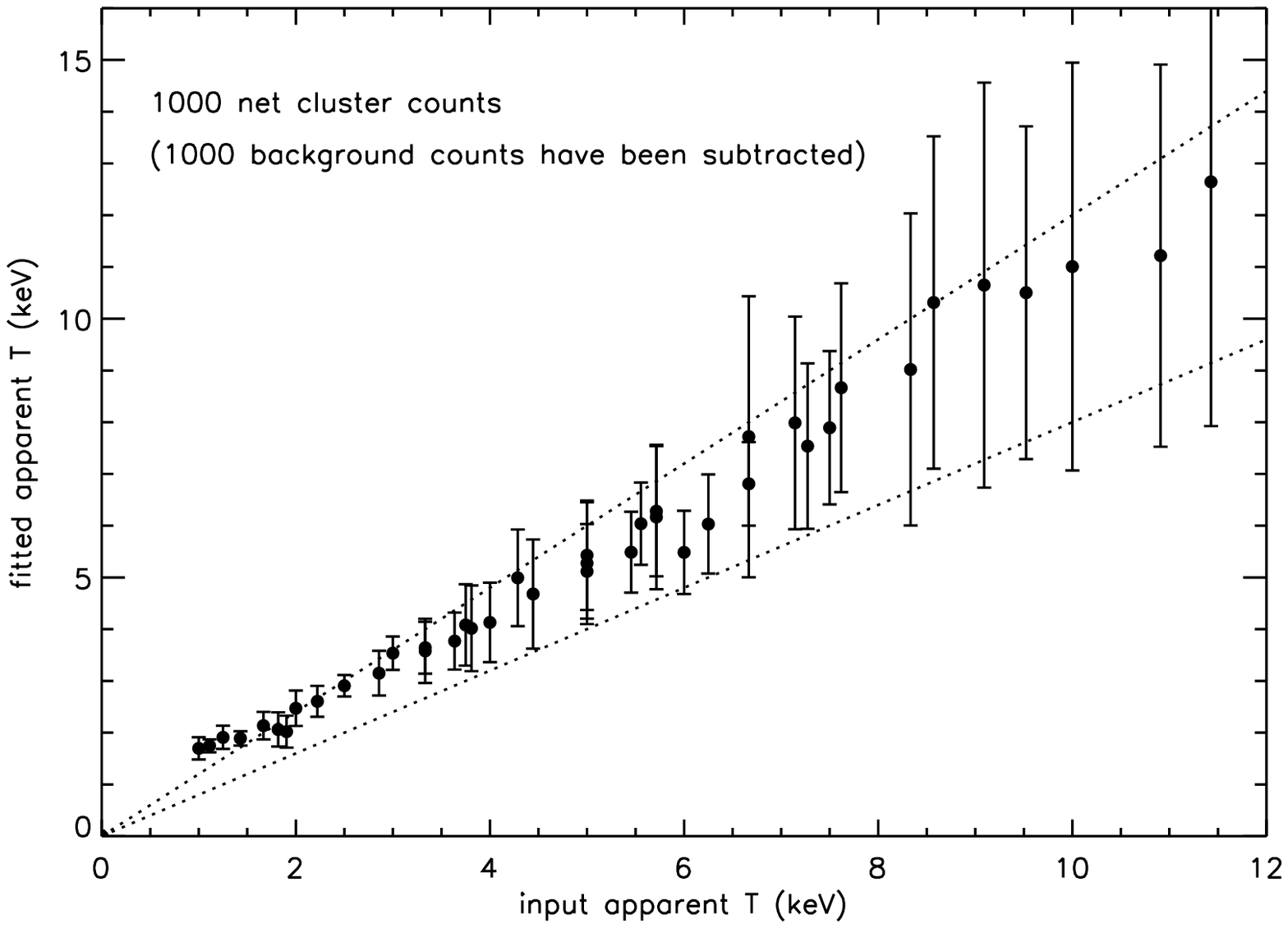}\\
\caption[Temps]{Fitted apparent temperature versus input apparent
temperature for simulated XMM cluster spectra after the subtraction of
100 (upper panel) and 1000 (lower panel) background counts. All input
spectra were created using {\tt fakeit} in {\sc xspec} and contained
1000 net cluster counts. The dotted lines delineate input apparent 
temperature plus and minus 20 per cent.}
\end{figure}

We estimated the likely temperature errors using {\sc xspec} simulations.  In
Figure~5, it can be seen that, for a wide range of input apparent temperatures,
the mean fitted apparent temperature is roughly equal to the input and that the
1 sigma errors bars fall within 20 per cent of the mean.  The {\sc xspec}-based
methodology used to produce this figure (i.e.~production of simulated spectra
from cluster and background spectral models, background subtraction, spectral
fitting etc.)  was similar to that used in Romer et al.~(1999) (e.g.~see Figure
4 and \S 5.2 of that paper) to estimate the errors on temperature fits for
clusters with known redshifts.  The only notable modification to that
methodology was the fixing of the redshift in the fits at $z=0$ rather than at
the redshift of the input spectrum.  As in Romer et al.~(1999), 20 fake spectra
per parameter combination were created and then fitted in order to derive a
stable mean and a realistic error distribution.  Figure~5 shows 84 parameter
combinations; six temperatures (2, 4, 6, 8, 10 and 12 keV), seven redshifts
($z=$ 0.05, 0.1, 0.2, 0.4, 0.6, 0.8, 1.0) and two background contaminations (100
and 1000 background counts for the upper and lower plots respectively).  The net
number of cluster counts after background subtraction was fixed at 1000; this
was the threshold adopted in Romer et al.~(1999) to predict the number of {\sc
Xcs} clusters that will yield temperature estimates.  For the purposes of
illustration we have shown two extremes of the expected background
contamination; most spectra will be contaminated by a few hundred background
counts (with an imposed upper limit of 1000, see \S 5.2 of Romer et al.~(1999)).
For typical clusters the true errors are likely to be less than 20 per cent,
because the number of cluster counts will be greater, but this assumption may be
more realistic for the distant, fainter clusters in which we are primarily
interested.  Unfortunately at present computational limitations prevent us from
carrying out fully self-consistent simulations where the expected cluster counts
are derived from the modelled cluster properties, and the XMM in-flight
background estimates are not yet available.

\begin{figure}
\centering
\leavevmode\epsfysize=5.3cm \epsfbox{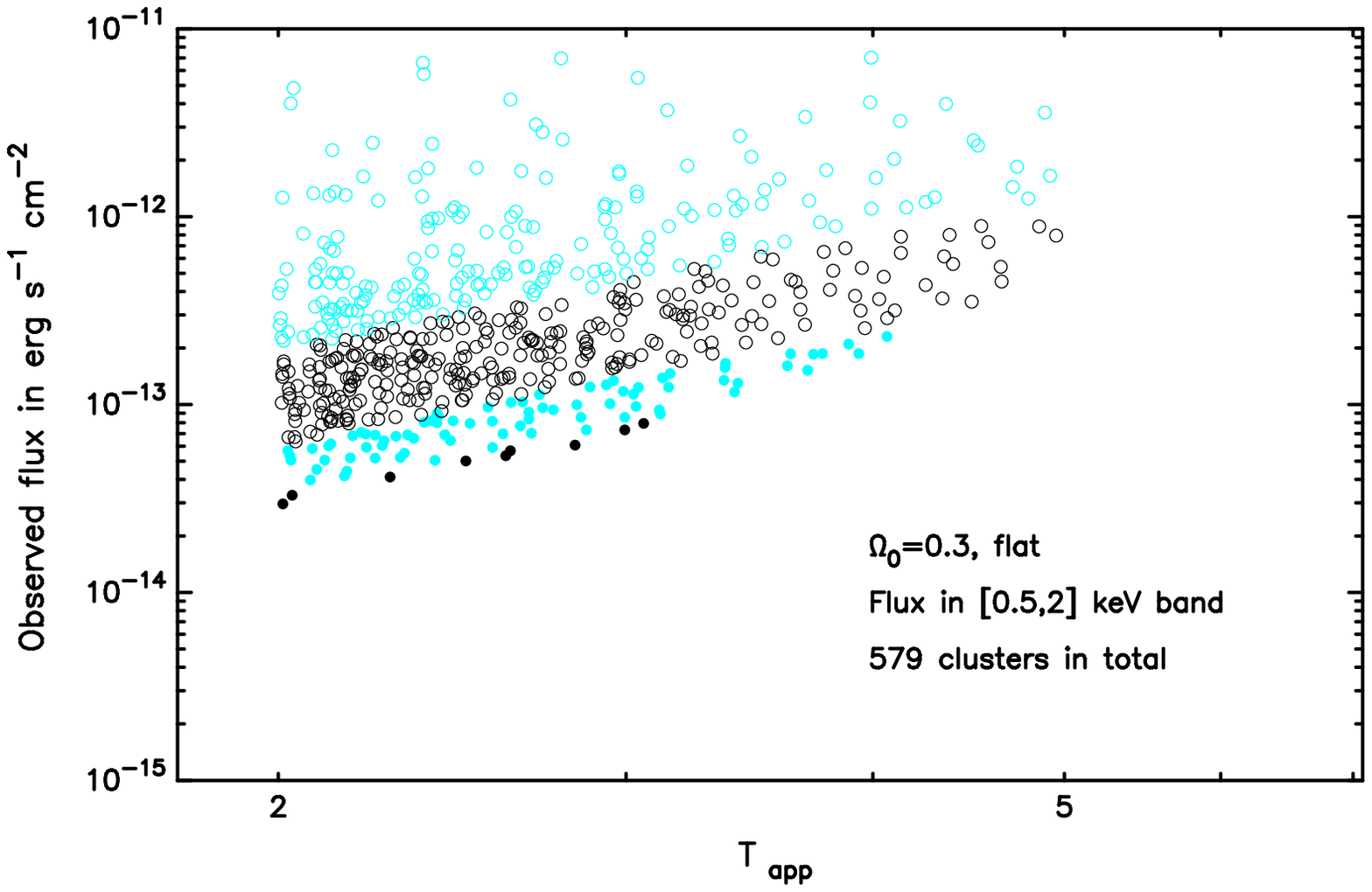}\\
\leavevmode\epsfysize=5.3cm \epsfbox{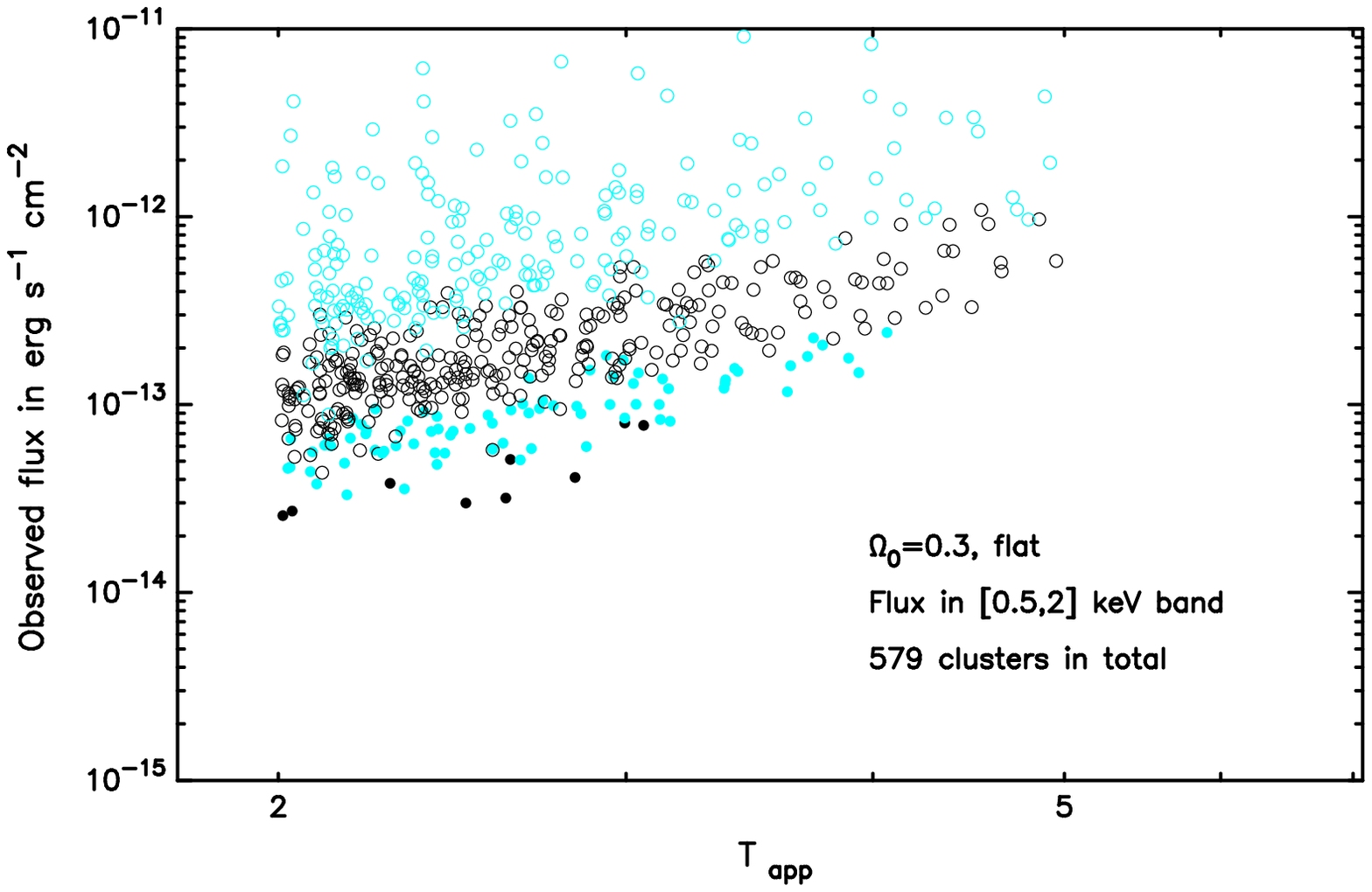}\\
\leavevmode\epsfysize=5.3cm \epsfbox{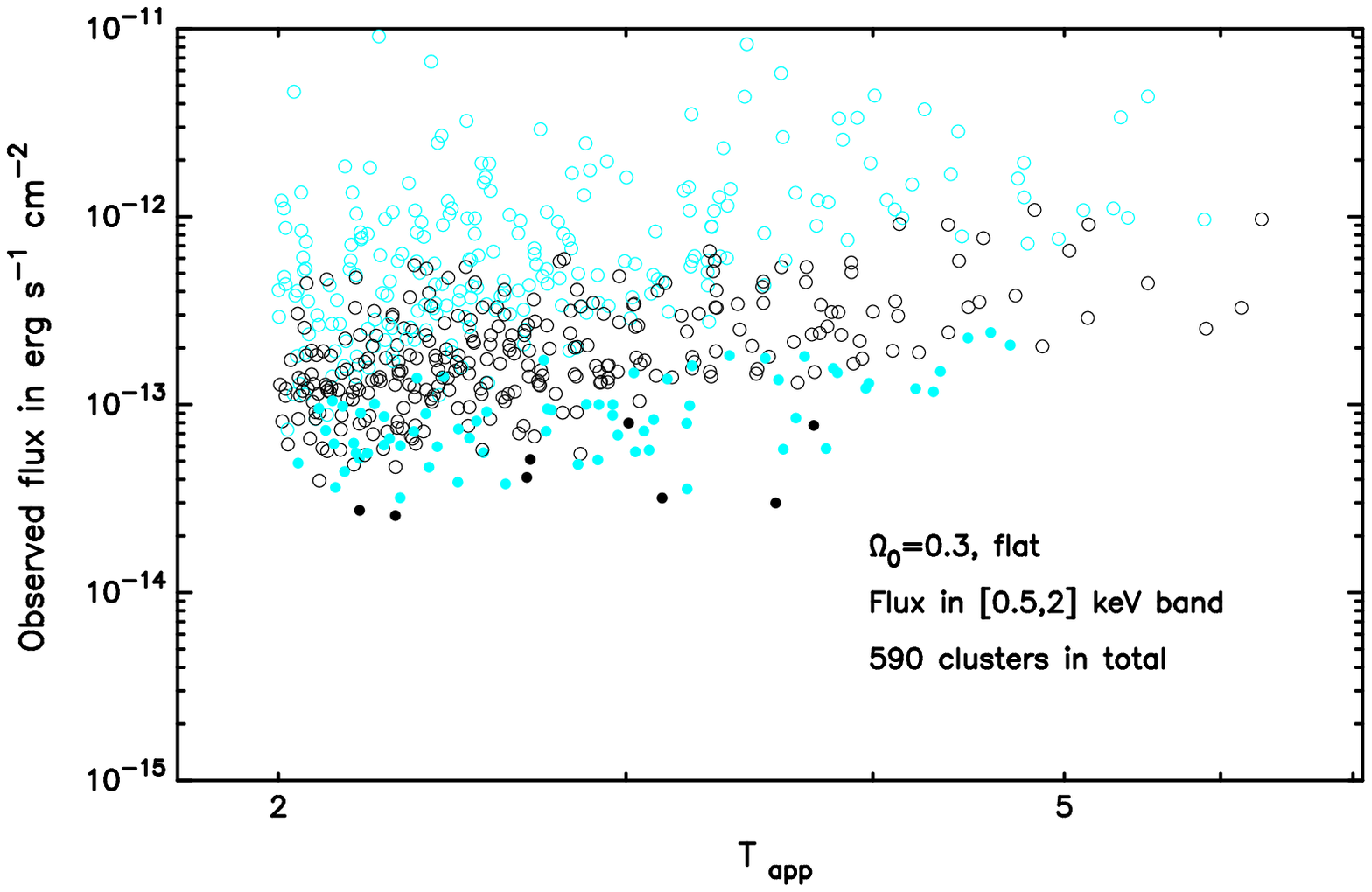}\\
\caption[scatter]{The expected location of clusters in the
X-ray flux versus apparent temperature plane. 
The clusters are divided
into four redshift intervals: $0 < z < 0.3$, cyan/grey open circles; $0.3 <
z < 0.6$, black open circles; $0.6 < z < 0.9$, cyan/grey points; $z
> 0.9$, black points. These simulations are for a flat
low-density Universe, for three years of {\sc Xcs} data. The upper panel assumes 
no error in the apparent temperatures, and no scatter in the 
luminosity--temperature relation. The middle panel introduces 
scatter in the luminosity (at fixed temperature), while the bottom panel is the 
realistic situation of both luminosity scatter and temperature measurement 
errors.}
\end{figure}

The results of the Monte Carlos are shown in Figure~6, in which the different
colours/symbols show different bins in redshift.  For reference, the upper panel
shows the location of clusters assuming the apparent temperatures are measured
precisely, and that there is no scatter in the luminosity at a given temperature
or errors in the temperature measurements.  The middle panel show the effect of
including luminosity scatter alone, which is not particularly significant, on
the same Monte Carlo realization.  The effect of adding temperature
uncertainties (in addition to luminosity scatter) is shown in the bottom panel
of Figure~6.  The plot shows only those clusters whose measured apparent
temperature exceeds 2 keV, regardless of what their true apparent temperature
is.  One could attempt to reduce the temperature errors by making extra pointed
observations once the clusters have been identified serendipitously, but the
high-redshift ones are typically only appearing in the longest serendipitous
exposures anyway.

A feature of the bottom panel is that it contains more clusters than the 
upper two panels. This is because while temperature errors in an individual 
cluster are as likely to be overestimates as underestimates, there are more 
low temperature clusters to `scatter up' than high temperature ones to scatter 
down. Careful Monte Carlo simulations are necessary to take out this 
bias (see e.g.~Viana \& Liddle 1999) when estimating the true temperature 
functions.

Returning to our motivation in making these plots, the aim was to test whether
the scattering of the points significantly decreases the ability to identify the
high-redshift cluster population within {\sc Xcs} using observed X-ray flux and
apparent temperature.  From Figure~6, we see that the segregation of clusters
with redshift remains quite good.  For example, drawing a line between the
points $(2,10^{-13})$ and $(7,10^{-12})$ clearly differentiates between local
clusters, with $z<0.3$, which lie above the line, and clusters with $z>0.6$
which lie below the line; it reduces the sample to 44\% of its original size 
without 
losing any of the $z>0.6$ clusters. A more stringent cut a factor of two lower 
in flux can create a sample of mostly high-redshift clusters; it reduces the 
sample to 117 clusters, including 65 of the 78 clusters above $z = 0.6$.

We conclude that the combination of apparent temperature and X-ray flux can be 
used as a redshift indicator.  Analysis of
the Monte Carlo data shows that some care will be needed; the general tendency
of the temperature errors to scatter clusters upwards in temperature usually
results in an overestimation of the redshift, though this can readily be
accounted for.  Taking scatter in luminosity and temperature into account, we
find that an estimated redshift with 33 per cent $1$-sigma uncertainty at high
redshift should be readily achievable, though the absolute scale relating
flux and apparent temperature to redshift will require calibration against
real data.  Most likely we have been quite conservative in the assessment of
temperature errors, so the actual accuracy of the estimation may considerably
exceed this in practice.

We should stress that Figure~6 is based on several assumptions, such as 
non-evolution of the luminosity--temperature relation with redshift, which are 
at best weakly tested by current observations. However, improved information 
from early XMM and Chandra observations can readily be incorporated when 
available to improve the use of flux and apparent temperature as a 
redshift estimator. 

A possible danger in using the flux is contamination of the cluster counts by
point sources, for example AGN within cluster galaxies.  For the ROSAT
satellite, with its low spatial resolution, this has proved a problem for
several clusters, especially those at high redshift; see for example Romer et
al.~(2000) and \S6.2.2 of Romer et al.~(1999).  The problem will be considerably
less significant for XMM given its much higher spatial resolution.

\subsection{X-ray spectral features}

For suitably luminous clusters one may see emission lines in the spectrum which
allows the degeneracy between $T$ and $z$ to be broken using the serendipitous 
X-ray data alone. The existence of such lines is illustrated in Figure~7, where 
we show model spectra for three clusters, chosen to have the same apparent 
temperature of $2$ keV and placed at redshifts $z = 0$, $0.5$ and $1$. 
The spectra were generated using the {\sc xspec} {\tt fakeit} routine from 
absorbed Raymond--Smith (1977) plasma models assuming a neutral hydrogen column 
density of $4 \times 10^{20}$ atoms/cm$^2$ and a metal abundance of 0.3 times 
the solar value. The spectra have been folded through the XMM EPIC-pn response 
files {\tt epn\_new\_rmf.fits} and {\tt epn\_thin\_arf.fits}, available from 
{\tt astro.estec.esa.nl}. For the purposes of this illustration, the exposure 
times were chosen so that each spectrum contained one million counts in the 
$0.5$--$10$ keV band when the metal abundance was zero, which is vastly higher 
than that expected for real clusters.

\begin{figure}
\centering
\leavevmode\epsfysize=5.6cm \epsfbox{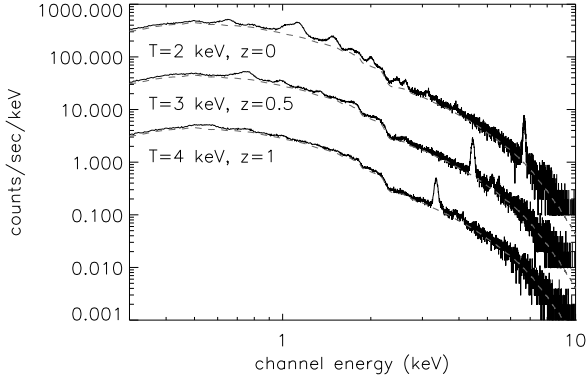}\\
\caption[Fig7]{Simulated spectra for clusters with apparent temperature $2$ keV, 
assuming an unrealistically large flux of a million counts for each 
cluster. The dashed lines show the zero metal abundance curves;
variations from these curves are due to the effects of line emission. To
allow easier comparison, the $T=3$ keV and $T=4$ keV spectra have been divided 
by 10 and 100 respectively before plotting.}
\end{figure}

In addition to the thermal bremsstrahlung continuum (the grey lines), several
features due to line emission are visible, for example the Iron K line complex
at 7 KeV and the rich collection of emission lines in the energy range 0.6 keV
to 2 keV, most notably the complex of iron, neon and magnesium lines at $\simeq
1$ keV.  As Figure~7 shows, the 7 keV (rest frame) Iron line complex moves to
lower energies as redshift increases.  A disadvantage of the Iron line complex
is that it falls at an energy at which both the cluster spectrum and energy
response of XMM has fallen off significantly.  In principle, one could use the
line complex at $\simeq 1$ keV, where XMM is significantly more sensitive,
though in practice this might only be possible for the low-temperature systems
as the contrast of the line emission against the thermal continuum drops off
rapidly with increasing temperature (see Figure~7).

Unfortunately, for the purposes of the {\sc Xcs}, the illustrative Figure~7 is 
unrealistic because it contains far too many counts and ignores the effect of 
the cosmic and particle backgrounds. Our guideline threshold for {\sc Xcs} 
clusters to have measurable 
temperatures is one thousand counts, and as seen from Figure~6 the brightest 
clusters (which are of course the nearby ones) will have up to around one 
hundred thousand counts (clusters of the same apparent temperature share the 
same spectral shape and so the count rate is proportional to the flux). In 
Figure~8 we show a more realistic, though still optimistic, simulation of a 
$z=1$, $T=4$ keV model spectrum containing ten thousand counts in the 
$0.5$--$10$ keV range, again ignoring the effects of background contamination. 
For spectra of this quality, or worse, one would need to use template fitting, 
rather than eyeballing, to derive redshift estimates. We conclude that X-ray 
lines may well allow accurate redshift estimation for many low-redshift 
clusters, but is unlikely to work well for high-redshift ones. Estimating the 
limiting redshift will require more detailed information on the backgrounds than 
is presently available.

\begin{figure}
\centering
\leavevmode\epsfysize=6cm \epsfbox{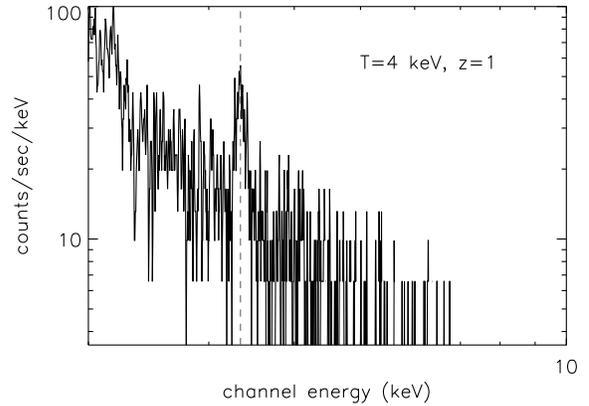}\\
\caption[Fig8]{Part of a cluster spectrum which contains ten thousand counts in 
the 0.5-10 keV band. This spectrum was derived from the same model that
produced the highest redshift of the three spectra presented in Figure~7.
The dotted grey line marks the expected position of the redshifted Iron K
line complex.}
\end{figure}

We note that few, if any, X-ray redshifts will be derived from the XMM RGS
cameras, despite them having much higher spectral resolution than the EPIC
cameras. This is because the RGS disperses energy from sources
across the field of view, so it would be extremely difficult to isolate the
spectrum from an {\sc Xcs} cluster (an extended off-axis source) from that
of the pointing target. For an example of how RGS can be used to study
nearby bright cluster targets, see Tamura et al.~(2000).

\section{Conclusions}

In this paper we have studied various aspects of apparent cluster temperatures 
relevant to XMM observations. The apparent temperature function will be readily 
derived from XMM data, but will primarily be useful in constraining the 
degenerate combination familiar from low-redshift cluster number density 
studies, $\sigma_8 \Omega_0^{-\alpha}$ with $\alpha \sim 0.5$, and is unlikely 
to usefully probe the density parameter itself. 

To fully exploit a galaxy cluster catalogue, cluster redshifts are essential, 
and we have studied how other X-ray properties can be combined with the apparent 
temperature to select follow-up candidates efficiently. Assuming point source 
contamination can be recognized from the high-resolution imaging, it appears 
that the cluster flux, when 
combined with the apparent temperature, will yield a good indication of the 
cluster redshift, 
especially once early XMM observations have been used to calibrate the relation.

Finally, we remark that further useful information 
towards estimating the redshifts may come from imaging data in the optical and 
infra-red once cluster candidates have been identified in the X-ray. A good 
example is the K band luminosity of the brightest cluster galaxy, which exhibits 
a tight correlation with redshift for 
X-ray luminous clusters, as shown by Collins \& Mann (1998) and Burke, Collins 
\& Mann (2000). In many cases it 
may also be possible to carry out 
photometric redshifting of cluster galaxies, for example using Sloan Digital Sky 
Survey and VISTA data.
 
\section*{ACKNOWLEDGMENTS}

We thank Jim Bartlett, Alain Blanchard and Joy Muanwong for useful 
discussions and Andy Ptak for advice on X-ray line features, and ARL and PTPV 
thank the Observatoire Midi-Pyr\'en\'ees for 
hospitality while part of this work was carried out.



\bsp
\end{document}